\documentclass[12pt]{article}
\newcounter{saveeqn}
\newcommand{\alpheqn}%
    {\setcounter{saveeqn}{\value{equation}}%
     \stepcounter{saveeqn}%
     \setcounter{equation}{0}%
     }
\newcommand{\reseteqn}%
    {\setcounter{equation}{\value{saveeqn}}%
     }
\begin{document}
\title{\normalsize \hfill UWThPh-2001-35 \\[1cm] \LARGE
Testing the Weisskopf--Wigner approximation 
by using neutral-meson--antimeson correlated states}
\author{G. V. Dass \\
\small Physics Department, Indian Institute of Technology \\
\small Powai, Mumbai 400076, India \\*[3.6mm]
W.\ Grimus \\
\small Institut f\"ur Theoretische Physik, Universit\"at Wien \\
\small Boltzmanngasse 5, A--1090 Wien, Austria}

\date{6 September 2001}

\maketitle

\begin{abstract}
We phenomenologically decompose the Weisskopf--Wigner approximation,
as applied to the neutral flavoured meson complexes, into three pieces and
propose tests for these pieces. Our tests hold for general decay
amplitudes and $M^0$--$\bar M^0$ mixing parameters. We concentrate on
C-odd $M^0 \bar M^0$ states and
stress the importance of such tests in view of the variety of
physics extracted from measurements on such complexes. Studying
the feasibility of the tests confines one
to the $K^0 \bar K^0$ system at present.
In particular, we show that the time dependence of the correlated decay
$\phi \to K^0 \bar K^0 \to 2 (\pi^+ \pi^-)$
is determined solely by the WWA and provides thus a clean test of it.

\vspace*{8mm}

\end{abstract}

\newpage

The complex formed by a neutral flavoured meson
$M^0$ (i.e., $K^0$, $D^0$, $B^0_d$, $B^0_s$) and its antiparticle $\bar M^0$
is an important environment for investigating (i) discrete
symmetries CP, T and CPT, (ii) quantum-mechanical correlations and also
(iii) new physics. The present phenomenology of this complex is commonly based
on the Weisskopf--Wigner approximation (WWA) \cite{wwa}. 
Since tiny effects are searched for in the $(M^0, \, \bar M^0)$ complex, it is
desirable to devise experimental tests of the WWA itself, 
independently of the physics searched for. One component of the WWA is the
exponential decay law 
(for theoretical studies see, e.g., Ref.~\cite{bernardini} and
papers cited therein), which has been tested \cite{norman} in
nuclear physics, and no deviations have been found, nor in particle
physics. Recently however, strong non-exponential
decay features have been observed in a quantum-mechanical system
\cite{fischer}. Ever since
the development of the WWA formalism for the $(K^0, \, \bar K^0)$ complex 
\cite{lee,kabir-book}, many studies have been performed to search for
possible deviations 
from the WWA; these would be important in the evaluation of experimental data 
on the $(M^0, \, \bar M^0)$ complex (for recent references see, e.g.,
\cite{wwa-k0k0bar}). In the present letter we study the evolution of the
correlated C-odd $M^0 \bar M^0$ state for testing the WWA. We
will phenomenologically split the WWA into three basic components and
investigate separately their consequences. 
We want to derive consequences of the WWA alone, independently of 
CP, T and CPT invariances (of mixing or of decay amplitudes)
and the $\Delta Q = \Delta F$ rule, where $F$ means
flavour (i.e., $F$ may be either $S$, $C$ or $B$). 

We first attempt a phenomenological anatomy of the WWA.
If we denote the general probability amplitudes for the transitions
$| M^0 \rangle \to | M^0 \rangle$,
$| M^0 \rangle \to |\bar M^0 \rangle$,
$| \bar M^0 \rangle \to | M^0 \rangle$ and 
$| \bar M^0 \rangle \to |\bar M^0 \rangle$,
respectively, as $a(t)$, $b(t)$, $\bar b(t)$ and $\bar a(t)$, 
where $t$ is the proper time, 
the WWA provides a model for these amplitudes.
We shall propose tests for the pieces of that model. In order to define these
pieces, we recall that the WWA introduces two independently propagating states
\begin{equation}\label{MHMLprop}
\left( \begin{array}{c} | M_H \rangle \\ | M_L \rangle 
\end{array} \right)
\stackrel{t}{\to} 
\left( \begin{array}{cc} \Theta_H(t) & 0 \\ 0 & \Theta_L(t) 
\end{array} \right)
\left( \begin{array}{c} | M_H \rangle \\ | M_L \rangle 
\end{array} \right)
\quad \mbox{with} \quad
\Theta_H(0) = \Theta_L(0) = 1 \,.
\end{equation}
The states\footnote{In the case of the neutral kaons, the long-lived
variety $K_L$ corresponds to the heavier state $M_H$, and the
short-lived $K_S$ corresponds to the lighter state $M_L$.}
$| M_{H,L} \rangle$ are suitable linear combinations of the two
flavour states $|M^0 \rangle$ and $|\bar M^0 \rangle$:
\begin{equation}\label{MHML}
\begin{array}{ll}
| M_H \rangle = p_H | M^0 \rangle + q_H | \bar M^0 \rangle \,,
& |p_H|^2 + |q_H|^2 = 1 \,, \\
| M_L \rangle = p_L | M^0 \rangle - q_L | \bar M^0 \rangle \,,
& |p_L|^2 + |q_L|^2 = 1 \,, 
\end{array}
\end{equation}
where $p_{H,L}$ and $q_{H,L}$ are complex constants. Because of the relevant
approximation, the 
[\nolinebreak\hspace{2pt}$(|M^0 \rangle$, $|\bar M^0 \rangle)$
$\leftrightarrow$ $(|M_H \rangle$, $| M_L \rangle)$\hspace{2pt}]
system is closed, the effect of the physical channels (like $\pi^+ \pi^-$) 
of 
$\stackrel{\scriptscriptstyle (-) \hphantom{0}}{M^0}$
decay being taken into account by the details of the propagation functions
$\Theta_{H,L}(t)$. Of course, the transformation (\ref{MHML}) is invertible:
\begin{equation}\label{inv}
\left( \begin{array}{c} | M^0 \rangle \\ | \bar M^0 \rangle 
\end{array} \right)
= A^{-1}
\left( \begin{array}{c} | M_H \rangle \\ | M_L \rangle 
\end{array} \right)
\quad \mbox{with} \quad
A = \left( \begin{array}{rr} p_H & q_H \\ p_L & -q_L 
\end{array} \right) \,.
\end{equation}
The ``decoupled'' propagation of $| M_{H,L} \rangle$ in Eq.~(\ref{MHMLprop})
restricts the otherwise general coefficients $a$, $b$, $\bar b$, $\bar a$.
Using (i) the closed nature of 
[\hspace{2pt}$(|M^0 \rangle$, $|\bar M^0 \rangle)$ $\leftrightarrow$
$(|M_H \rangle$, $| M_L \rangle)$\hspace{2pt}],
(ii) Eqs.~(\ref{MHML}) and (\ref{inv}) and (iii) the definitions for
the coefficients $a$, $b$, $\bar b$, $\bar a$, one gets
\begin{eqnarray}
&& \Omega 
\left( \begin{array}{c} | M_H \rangle \\ | M_L \rangle 
\end{array} \right) =
\Omega\, A
\left( \begin{array}{c} | M^0 \rangle \\ | \bar M^0 \rangle 
\end{array} \right) =
A\, \Omega 
\left( \begin{array}{c} | M^0 \rangle \\ | \bar M^0 \rangle 
\end{array} \right) = \nonumber \\ 
&&
= A \left(
\begin{array}{cc} a & b \\ \bar b & \bar a \end{array}
\right)
\left( \begin{array}{c} | M^0 \rangle \\ | \bar M^0 \rangle 
\end{array} \right) = 
A \left(
\begin{array}{cc} a & b \\ \bar b & \bar a \end{array}
\right) A^{-1} 
\left( \begin{array}{c} | M_H \rangle \\ | M_L \rangle 
\end{array} \right) \,, \label{omega}
\end{eqnarray}
where $\Omega$ denotes the time-development operator.
Using Eq.~(\ref{MHMLprop}), this means
\begin{equation}\label{comparison}
\left( 
\begin{array}{cc} \Theta_H & 0 \\ 0 & \Theta_L \end{array}
\right) =
A \left(
\begin{array}{cc} a & b \\ \bar b & \bar a \end{array}
\right) A^{-1} \,.
\end{equation}
Comparing the diagonal elements, one obtains
\begin{equation}\label{diag}
\Theta_H = \frac{1}{2} ( a + \bar a) + \gamma
\quad \mbox{and} \quad
\Theta_L = \frac{1}{2} ( a + \bar a) - \gamma \,,
\end{equation}
with
\begin{equation}\label{gamma,D}
\gamma = \left\{ \frac{1}{2} ( a - \bar a ) ( p_H q_L - q_H p_L ) +
b p_H p_L + \bar b q_H q_L \right\}/D
\quad \mbox{and} \quad
D = p_H q_L + p_L q_H \,.
\end{equation}
Computing the off-diagonal elements (12 and 21 elements) of
Eq.~(\ref{comparison}), similarly gives
\begin{equation}
\begin{array}{c}
\Delta_{12} = \left\{ (a - \bar a) p_H q_H - b p_H^2 + \bar b q_H^2
\right\}/D \\
\Delta_{21} = \left\{ (a - \bar a) p_L q_L + b p_L^2 - \bar b q_L^2
\right\}/D \,.
\end{array}
\end{equation}
The vanishing of the off-diagonal elements $\Delta_{12}$ and
$\Delta_{21}$ is the lack of ``vacuum regeneration'' (viz.\ the
absence of 
$| M_{H,L} \rangle \to | M_{L,H} \rangle$ transitions) in the
WWA. This directly gives the first piece of the WWA:
\begin{equation}\label{wwa1}
\mbox{WWA1:} \quad \bar a(t) - a(t) = \beta b(t) \,, \quad
\bar b(t) = \alpha b(t) \,,
\end{equation}
where 
\begin{equation}\label{alphabeta}
\alpha = \frac{p_H p_L}{q_H q_L} 
\quad \mbox{and} \quad
\beta = \frac{p_L}{q_L} - \frac{p_H}{q_H} \,.
\end{equation}
Of course, Eq.~(\ref{wwa1}) holds for all $t$ and any
$\Theta_{H,L}(t)$. The second piece of the WWA arises from the
diagonal elements of Eq.~(\ref{comparison}). Using Eq.~(\ref{diag}),
one obtains the $t$-dependence of the coefficients $a$, $b$, $\bar b$,
$\bar a$ as
\alpheqn
\begin{eqnarray}
& \mbox{WWA2:} \quad &
a + \bar a = \Theta_H + \Theta_L \,,     \label{wwa2a} \\ &&
b = q_H q_L (\Theta_H - \Theta_L)/D \,,  \label{wwa2b}
\end{eqnarray}
\reseteqn
wherein $\gamma$ has been simplified with the help of Eq.~(\ref{wwa1}).
While Eq.~(\ref{wwa2b}) is expressed in terms of $b$, one could have
equivalently written $\bar a - a$ or $\bar b$ in terms of 
$\Theta_H - \Theta_L$, because of Eq.~(\ref{wwa1}). Thus all the four
coefficients $a$, $b$, $\bar b$, $\bar a$ are given in terms of
the functions $\Theta_{H,L}$ which are so far not specified. One may
note that Eq.~(\ref{wwa2a}) is expected because it is merely the
invariance of the trace under the similarity transformation expressed by
Eq.~(\ref{comparison}). The third piece of the WWA is the
specification of $\Theta_{H,L}$ in terms of the exponential law:
\begin{equation}\label{wwa3}
\mbox{WWA3:} \quad 
\Theta_{H,L}(t) = \exp (-it \lambda_{H,L}) 
\quad \mbox{with} \quad
\lambda_{H,L} = m_{H,L} - \frac{i}{2} \Gamma_{H,L} \,,
\end{equation}
where, as usual, $m_{H,L}$ are the real masses and
$\Gamma_{H,L}$ the real decay widths of $M_{H,L}$.
With all the pieces of the WWA put in, one eventually arrives at
(for a convenient summary, see, e.g., \cite{branco,sanda})
\begin{equation}
\label{ab}
\begin{array}{lll}
a(t) & = & g_+(t) - \theta g_-(t) \, , \\*[2mm]
b(t) & = & {\displaystyle \frac{q}{p}}\, \sqrt{1-\theta^2} g_-(t) \, , \\*[2mm]
\bar a(t) & = & g_+(t) + \theta g_-(t) \, , \\*[2mm]
\bar b(t) & = & {\displaystyle \frac{p}{q}}\, \sqrt{1-\theta^2} g_-(t) \,,
\end{array}
\end{equation}
where
\begin{eqnarray}
&& g_\pm(t) = \frac{1}{2} \left\{
\exp (-it \lambda_H) \pm \exp (-it \lambda_L) \right\} \,, \\
&& \frac{q}{p}= \sqrt{\frac{q_H q_L}{p_H p_L}}\,, \quad
\theta = \frac{q_H/p_H - q_L/p_L}{q_H/p_H + q_L/p_L} \,, \\
&& \alpha = \left( \frac{p}{q} \right)^2 \,, \quad
\beta = 2\, \frac{p}{q} \frac{\theta}{\sqrt{1-\theta^2}} \,.
\end{eqnarray}
Note that $\theta$, but not $q/p$, is rephasing-invariant. Thus
both the real and the imaginary parts of $\theta$
are in principle measurable, and also $|q/p|$, but not the phase of
$q/p$ \cite{branco,lavoura}.
A non-zero $\theta$ signifies CPT and CP non-invariance in mixing; 
similarly, a non-zero $|q/p|-1$ signifies T and CP non-invariance.
In the usual explicit calculations based on the WWA, the full model of
Eq.~(\ref{ab}) is used. Our interest, in contrast, is in the three
ingredients (WWA1, WWA2, WWA3) and testing them; the tests of WWA2
(without WWA3) are bound to be only qualitative because 
they can merely examine the differences
between the unknown $\Theta_H$ and $\Theta_L$.

We now consider decays of the correlated 
$M^0 \bar M^0$ states
\begin{equation}
\label{psi}
| \psi_\epsilon \rangle = {\textstyle \frac{1}{\sqrt{2}}}
\left[
| M^0 ( \vec k ) \rangle \otimes | \bar M^0 ( - \vec k ) \rangle +
\epsilon
| \bar M^0 ( \vec k ) \rangle \otimes | M^0 ( - \vec k ) \rangle
\right] \,,
\end{equation}
where $\epsilon$ denotes the charge conjugation value;
$\epsilon = +1$ for the C-even case,  
$\epsilon = -1$ for the C-odd case.  
If one detects the decay channel $f$ at time $t_\ell$ and the
channel $g$ at time $t_r$, the decay rate of $| \psi_\epsilon \rangle$
is 
\begin{eqnarray}
\lefteqn{R_\epsilon(f,t_\ell;g,t_r) =} \nonumber \\
&& \frac{1}{2} \bigg| 
(a_\ell \bar b_r + \epsilon \bar b_\ell a_r ) A_f A_g + 
(b_\ell \bar a_r + \epsilon \bar a_\ell b_r ) \bar A_f \bar A_g + 
\nonumber \\
&&
\left( a_\ell \bar a_r + b_\ell \bar b_r + 
\epsilon ( \bar a_\ell a_r + \bar b_\ell b_r ) \right) \,
\frac{1}{2} (A_f \bar A_g + \bar A_f A_g) + 
\nonumber \\
&& 
\left( a_\ell \bar a_r - b_\ell \bar b_r - 
\epsilon ( \bar a_\ell a_r - \bar b_\ell b_r ) \right) \,
\frac{1}{2} (A_f \bar A_g - \bar A_f A_g) \bigg|^2 \,.
\label{wwa0}
\end{eqnarray}
In this expression the transition amplitudes are defined as
\begin{equation}\label{amp}
\langle f | T |      M^0 \rangle =      A_f \,, \quad
\langle f | T | \bar M^0 \rangle = \bar A_f \,, \quad
\langle g | T |      M^0 \rangle =      A_g \,, \quad
\langle g | T | \bar M^0 \rangle = \bar A_g \,,
\end{equation}
and
\begin{equation}
\stackrel{\scriptscriptstyle (-)}{a}_{\hskip-2pt \ell}\, \equiv 
\stackrel{\scriptscriptstyle (-)}{a}\!(t_\ell) \,, \quad
\stackrel{\scriptscriptstyle (-)}{a}_{\hskip-2pt r}\, \equiv 
\stackrel{\scriptscriptstyle (-)}{a}\!(t_r) \,, \quad
\stackrel{\scriptscriptstyle (-)}{b}_{\hskip-2pt \ell}\, \equiv 
\stackrel{\scriptscriptstyle (-)}{b}\!(t_\ell) \,, \quad
\stackrel{\scriptscriptstyle (-)}{b}_{\hskip-2pt r}\, \equiv 
\stackrel{\scriptscriptstyle (-)}{b}\!(t_r) \,.
\end{equation}
The form of Eq.~(\ref{wwa0}) assumes only that aspect of the WWA which
was mentioned immediately after Eq.~(\ref{MHML}). We shall consider the
consequences of WWA1, WWA2, WWA3, successively, for the coefficients 
$a$, $b$, $\bar b$, $\bar a$. The aim is a comparison with experiment.

For the following we concentrate on the decay of $| \psi_- \rangle$, i.e., 
the C-odd case. Use of WWA1 gives
\begin{eqnarray}
\lefteqn{R_-(f,t_\ell;g,t_r) =} \nonumber \\
&& \frac{1}{2} \left| 
(a_\ell b_r - b_\ell a_r ) \left( 
\alpha A_f A_g - \bar A_f \bar A_g + 
\beta \, \frac{1}{2} (A_f \bar A_g + \bar A_f A_g) \right) +
\right. \nonumber \\
&& \left.
\left( 2 a_\ell a_r - 2 \alpha b_\ell b_r + \beta 
( a_\ell b_r + b_\ell a_r ) \right) \,
\frac{1}{2} (A_f \bar A_g - \bar A_f A_g) \right|^2 \,.
\label{R+wwa1}
\end{eqnarray}
This is the general form of the decay rate where Eq.~(\ref{wwa1}) has
been used.

The case that 
\begin{equation}\label{cond}
A_f \bar A_g - \bar A_f A_g = 0 
\end{equation}
deserves particular attention,
because then the time dependence is ``factored out'':
\begin{equation}\label{Rfact}
R_-(f,t_\ell;g,t_r) = \frac{1}{2} 
\left| a_\ell b_r - b_\ell a_r \right|^2 \, \left|
\alpha A_f A_g - \bar A_f \bar A_g + 
\beta A_f \bar A_g \right|^2 \,.
\end{equation}
This relation is a powerful test of the property of lack of vacuum
regeneration in the WWA: for all decay channels satisfying
Eq.~(\ref{cond}), the $(t_\ell \leftrightarrow t_r)$-symmetric time
dependence must be the same for a given choice of $M^0$. For
Eq.~(\ref{cond}) to hold, the decay amplitudes for the channels $f$ and
$g$ must be completely specified, using not only the particle content
of $f$ and $g$, but also their configurations of spins and
momenta. One is, therefore, led to consider spinless decay products
with a situation wherein there is no variable Lorentz scalar. Thus
spinless two-body channels (e.g., $\pi^+ \pi^-$) and effective
two-body channels seem interesting. (One possibility is the $3\pi$
mode where one pion, say $\pi_1$, moves back-to-back with the
remaining two, with no relative momentum between $\pi_2$ and $\pi_3$;
another possibility arises when
$\pi_1$ is created at rest.\footnote{These configurations have non-zero
amplitudes in the standard theory of $K \to 3\pi$;
see, e.g., Ref.~\cite{paver}. 
We thank H. Neufeld for discussions on this point.}) 
With this in
mind one may write $f=g$ in Eq.~(\ref{Rfact}).

For considering WWA2, Eqs.~(\ref{wwa2a}) and (\ref{wwa2b}),
we use also the first relation of
Eq.~(\ref{wwa1}) to obtain, apart from an overall constant,
\begin{eqnarray}
a_\ell b_r - b_\ell a_r & \to & 
(\Theta_H + \Theta_L)_\ell (\Theta_H - \Theta_L)_r -
(\Theta_H - \Theta_L)_\ell (\Theta_H + \Theta_L)_r =
\nonumber \\ &&
2\, \left( 
\Theta_L(t_\ell) \Theta_H(t_r) -  \Theta_H(t_\ell) \Theta_L(t_r)
\right) \,. \label{Rfact-timedep}
\end{eqnarray}
Due to WWA2, therefore, the unknown time dependence of
Eq.~(\ref{Rfact}) is now determined by the characteristics
$\Theta_{H,L}(t)$ of the WWA. The greater the difference between
$\Theta_H(t)$ and $\Theta_L(t)$, the more pronounced this time
dependence would be because $\Theta_H - \Theta_L$ occurs linearly in
Eq.~(\ref{Rfact-timedep}). Unfortunately, this ``test'' of the WWA
cannot be quantified because the $\Theta_{H,L}$ are yet unknown.

If one introduces the exponential law, viz.\ WWA3, the above feature
comes to the surface\footnote{This result can be shown to be contained
in the explicit calculation of Ref.~\cite{xing}, wherein the relevant
result, Eq.~(4), was obtained by making the assumption of CPT invariance, 
viz.\ $\beta = 0$. Our derivation is based on simpler and more general 
considerations. \label{ftnlabel}}:
\begin{equation}\label{timedep}
\left| a_\ell b_r - b_\ell a_r \right|^2 \to
e^{-\Gamma t_+}
\left\{ \cosh \left( {\textstyle \frac{1}{2}} \Delta \Gamma t_- \right)
- \cos \left( \Delta m\, t_- \right) \right\} \,,
\end{equation}
where we have used the definitions
$\Gamma = (\Gamma_H + \Gamma_L)/2$, 
$\Delta \Gamma = \Gamma_H - \Gamma_L$,
$\Delta m = m_H - m_L$ and $t_\pm = t_\ell \pm t_r$.
It is worth remarking that the time dependences (\ref{Rfact}),
(\ref{Rfact-timedep}) and (\ref{timedep}) for
$R_-(f,t_\ell;f,t_r)$ do not depend on any assumptions about the decay
amplitudes 
${\stackrel{\scriptscriptstyle (-)}{A}}_f$ -- in particular,
their behaviour under CP, T and CPT transformations and the
$\Delta Q = \Delta F$ rule; similarly, the constants $p_{H,L}$,
$q_{H,L}$ have been kept general; CP, T and CPT non-invariances
($\beta \neq 0$, $|\alpha| \neq 1$) have been allowed throughout.
The time dependence (\ref{timedep}) of $R_-(f,t_\ell;f,t_r)$ 
is, therefore, the test of the full WWA; it is a 
specific version of the general $(t_\ell \leftrightarrow t_r)$-symmetric
form in Eq.~(\ref{Rfact}). Note that so far the WWA has been used in
its general form, the exact values of the constants $\alpha$ and
$\beta$ appearing in Eq.~(\ref{alphabeta}) have not been exploited.
A side remark: the vanishing of Eqs.~(\ref{Rfact}),
(\ref{Rfact-timedep}) and (\ref{timedep}) for $t_\ell = t_r$ is merely
the quantum-mechanical expectation that
$R_-(f,t;f,t)$ vanishes (see, e.g., Ref.~\cite{enz}).
This feature is already present in Eq.~(\ref{wwa0})
with $f=g$, $t_\ell = t_r$ and $\epsilon = -1$;
it does not require WWA1, WWA2 and WWA3.

Another test of the general WWA framework of Eq.~(\ref{wwa0}) arises
by choosing $\epsilon = -1$ and $t_\ell = t_r$ for any 
$f \neq g$. Now, in complete contrast to Eq.~(\ref{cond}), only the
amplitude combination $A_f \bar A_g - \bar A_f A_g$ contributes to the
rate. Once again, one has a factorization of the time dependence:
\begin{equation}\label{tl=tr}
R_-(f,t_\ell;g,t_\ell) = \frac{1}{2}
\left| a_\ell \bar a_\ell - b_\ell \bar b_\ell \right|^2 \,
\left| A_f \bar A_g - \bar A_f A_g \right|^2 \,,
\end{equation}
leading to channel independence of the time dependence of $R_-$. 
If one now uses
WWA1 + WWA2, along with the values of $\alpha$ and $\beta$ from
Eq.~(\ref{alphabeta}), the time dependence becomes
\begin{equation}\label{test3}
R_-(f,t_\ell;g,t_\ell) \to 
\left| \Theta_H(t_\ell) \Theta_L(t_\ell) \right|^2 =
\frac{1}{16} \left[ ( |\Theta_H(t_\ell)| + |\Theta_L(t_\ell)| )^2 -
( |\Theta_H(t_\ell)| - |\Theta_L(t_\ell)| )^2 \right]^2 \,.
\end{equation}
Here, the difference between $\Theta_H$ and $\Theta_L$ makes only an
additive contribution; its role is therefore not as important as in
Eq.~(\ref{Rfact-timedep}). 
With WWA3, one gets\footnote{Here, footnote \ref{ftnlabel} applies as well.} 
\begin{equation}\label{test3a}
R_-(f,t_\ell;g,t_\ell) \to e^{-2\Gamma t_\ell} 
\end{equation}
as the channel-independent time dependence for a given choice of
$M^0$, which tests the full WWA.

We now consider the feasibility of our WWA tests. For the tests
in Eqs.~(\ref{Rfact}), (\ref{Rfact-timedep}), (\ref{timedep}), one
needs spinless two-body (or, effectively two-body) channels. For 
$M^0 = B^0_{d,s}$, the relevant branching ratios are very small. For 
$M^0 = D^0$, the relevant $| \psi_- \rangle$ states have not yet been
well-studied. This leads to $M^0 = K^0$. 
The obvious choice would then be to compare 
$\pi^+ \pi^-$ with $\pi^0 \pi^0$ for $f=g$ as a test of the channel
independence of the correlated decay rate as a function of $t_\ell$
and $t_r$. In this case, the decay amplitudes 
${\stackrel{\scriptscriptstyle (-)}{A}}_{f,g}$
are reasonably well studied. However, even without the WWA1, viz.\
already in Eq.~(\ref{wwa0}), the time dependence is expected to be
very nearly the same for these final states, as will be shown in the next
paragraph. One is, therefore, led to compare the two choices $\pi \pi$ and 
special cases of $\pi \pi \pi$ for $f=g$ and $M^0 = K^0$. For further choices,
one has to wait for future data.

We now give the detailed reason why the choices 
$\pi^+ \pi^-$ and $\pi^0 \pi^0$ for $f=g$ and $M^0 = K^0$
are not practically useful for testing the channel independence of the time
dependence of the rate (\ref{Rfact}). Considering the rate (\ref{wwa0}),
where WWA1 has \emph{not} yet been used, we observe that the ratios 
$A_f A_f : \bar A_f \bar A_f : A_f \bar A_f$ for $f=g$ determine its time
dependence. We will show that these ratios are the same for 
$f = \pi^+ \pi^-$ and $f = \pi^0 \pi^0$, except for quantities of second
order of smallness, i.e., of the order of the CP-violating quantity
$\varepsilon'$ which denotes CP violation in the decay amplitudes
\cite{branco}. In detail, let us write \cite{wolfenstein} 
\alpheqn
\begin{eqnarray}
&& A_I = \langle I | T | K^0 \rangle = a_I\,  e^{i\delta_I}
(1 + i\varphi_I + \beta_I + i\alpha_I ) \\ 
&& \bar A_I = \langle I | T | \bar K^0 \rangle = a_I\,  e^{i\delta_I}
(1 - i\varphi_I - \beta_I + i\alpha_I ) \,,
\end{eqnarray}
\reseteqn
for the decay amplitudes to the two isospin states $I = 0,\, 2$ of the 
$\pi \pi$ system; $\varphi_I$, $\beta_I$, $\alpha_I$  are (supposedly small)
real parameters expressing CP and T non-invariance, CP and CPT non-invariance,
T and CPT non-invariance, respectively; the $a_I$ are real; the $\delta_I$ are
the $\pi \pi$ scattering phase shifts at the c.m.\ energy which equals the
kaon mass. The isospin value $I$ appears as subscript on various
quantities. Then, with obvious meaning of the subscripts $+-$ and $00$, 
one obtains 
\alpheqn
\begin{eqnarray}
\lefteqn{A_{+-} A_{+-} : \bar A_{+-} \bar A_{+-} : A_{+-} \bar A_{+-}
=} \nonumber \\ &&
1 : 
[1 - 2\sqrt{2} \omega \zeta - 4\sigma + 4\sigma (2\sigma + i\alpha_0)] :
[1 -  \sqrt{2} \omega \zeta - 2\sigma + 2\sigma (\sigma + i\alpha_0)] \,, 
\label{ratios+-} \\
\lefteqn{A_{00} A_{00} : \bar A_{00} \bar A_{00} : A_{00} \bar A_{00}
=} \nonumber \\ &&
1 :
[1 + 4\sqrt{2} \omega \zeta - 4\sigma + 4\sigma (2\sigma + i\alpha_0)] :
[1 + 2\sqrt{2} \omega \zeta - 2\sigma + 2\sigma (\sigma + i\alpha_0)] \,,
\label{ratios00}
\end{eqnarray}
\reseteqn
where we have used the notation
\begin{equation}
\omega = \frac{a_2}{a_0} e^{i(\delta_2-\delta_0)} \,, \quad
\sigma = \beta_0 + i\varphi_0 \,, \quad
\zeta = i(\varphi_2 - \varphi_0) + (\beta_2 - \beta_0) +
        i(\alpha_2  - \alpha_0 ) \,.
\end{equation}
Here, $\omega$ is small because of the $\Delta I = 1/2$ rule; we
have retained small quantities up to only second order. Note that the only
differences in the ratios in Eqs.~(\ref{ratios+-}) and
(\ref{ratios00}) are due 
to the $\omega \zeta$ terms which are of the $\varepsilon'$ type: they are
proportional to $a_2$ and to a combination of $\varphi_I$, $\beta_I$ and
$\alpha_I$. Usually, one takes $\beta_I = \alpha_I = 0$; 
then, $\omega \zeta$ is directly seen to be $\sqrt{2}\, \varepsilon'$.
In the ratios in Eqs.~(\ref{ratios+-}) and (\ref{ratios00}),
quantities of first (viz.\ $\sigma$) and zeroth (viz.\ 1) order of smallness
are also present, apart from other quantities of second order (viz.\
$\sigma^2$, $\sigma \alpha_0$). Thus the ratios (\ref{ratios+-}) and
(\ref{ratios00}) are the same to a very good approximation. The observability
of a difference in the time distributions for the $+-$ and $00$ channels would
require, therefore, very accurate data, in general, even without the WWA1.

Let us consider the feasibility of the test of WWA3 with $M^0 = K^0$ and
$f = g = \pi^+ \pi^-$, for which the corresponding rate is being measured at
DA$\Phi$NE \cite{eberhard}. 
To get an estimate of the magnitude of the rate (\ref{Rfact}),
we neglect the CPT-violation parameter $\beta$ and the parameter
$\varepsilon'$. Also, we retain small quantities to only the lowest
contributing order. In this way we obtain
\begin{equation}\label{result}
R_-(f,t_\ell;f,t_r) \simeq \left( \Gamma(K_S \to \pi^+ \pi^-) \right)^2
\left| \eta_{+-} \right|^2 
e^{-\Gamma t_+}
\left\{ \cosh \frac{1}{2} \Delta \Gamma t_- - \cos \Delta m t_- \right\} \,,
\end{equation} 
where $\eta_{+-}$ denotes the ratio of the decay amplitudes of $K_L$
and $K_S$ into $\pi^+ \pi^-$.
DA$\Phi$NE will produce an adequate number of $K^0 \bar K^0$ pairs so
as to overcome the CP suppression in the rate (\ref{result}) \cite{eberhard}; 
its \emph{time dependence} is a clear
consequence of the full WWA, irrespective of any T, CP and CPT
violations or the validity of the $\Delta S = \Delta Q$ rule, as noted
immediately following Eq.~(\ref{timedep}).

Now we come to the feasibility of the tests (\ref{tl=tr}), (\ref{test3}),
(\ref{test3a}). The relevant time variable is only $t_+$ because 
$t_- = t_\ell - t_r$ vanishes now. But the variable $t_+$ is difficult to
measure at the present asymmetric B factories \cite{BELLE-BABAR} (see, e.g.,
also Ref. \cite{aleksan}). 
One is thus led to the choice $M^0 = K^0$ again. Since
now $f \neq g$, one can consider many possible choices for $f$ and $g$.

In summary, we have defined/derived the three pieces of the WWA from
the point of view of phenomenological applications. We have then
proposed tests for these pieces in a general way, without making any
assumptions about the decay amplitudes
${\stackrel{\scriptscriptstyle (-)}{A}}_{f,g}$
of 
$\stackrel{\scriptscriptstyle (-)\hphantom{0}}{M^0}$ decay
and about the constants $p_{H,L}$, $q_{H,L}$ of Eq.~(\ref{MHML}).
Our overall framework of Eq.~(\ref{wwa0}) assumes the WWA; our purpose was
to see the successive consequences of the three pieces of the WWA. The
tests (\ref{Rfact}), (\ref{Rfact-timedep}), (\ref{timedep}) involve
checking the channel independence of the observed rates, successively
for WWA1, WWA2 and WWA3. The last of these, Eq.~(\ref{timedep}), tests
also the exponential decay law for any choice of $f=g$ and $M^0$. 
The same holds for the three tests
(\ref{tl=tr}), (\ref{test3}), (\ref{test3a}). All these, at present,
are feasible for the choice $M^0 = K^0$. The first set, viz.\
(\ref{Rfact}), (\ref{Rfact-timedep}), (\ref{timedep}), is further
restricted to the comparison of the choices $\pi \pi$ and special
cases of $\pi \pi \pi$ for $f=g$. Hopefully, such tests can be
performed soon. For the other choices of $M^0$, one has to wait for
the future.

\end{document}